\documentclass[epsfig,onecolumn]{mn2e}
\usepackage{epsfig}
\usepackage{graphicx,amssymb}
\def\beq#1{\begin{equation}\label{#1}}
\def\eeq{\end{equation}}
\def\beqa#1{\begin{eqnarray}\label{#1}}
\def\eeqa{\end{eqnarray}}

\def\epeiso{$E_{\rm p,i}$ -- $E_{\rm iso}$}

\def\sax{{\it Beppo}SAX }
\def\swift{{\it Swift}}
\def\fermi{{\it Fermi}}

\def\h0{H$_{\rm 0}$~}

\title[The $E_{\rm p,i}$ -- $E_{\rm iso}$ relation]{Standardizing the GRBs with  the $\mathbf{Amati}$ $E_{\rm p,i}$ -- $E_{\rm iso}$  relation: the updated Hubble diagram and implications for cosmography } \author[M.Demianski et al.]{Marek Demianski$^{1,2}$ and Ester Piedipalumbo$^{3,4}$\\
$^1$ Institute for Theoretical Physics,
University of Warsaw,  Hoza 69, 00-681 Warsaw, Poland  \\
$^2$ Department of Astronomy, Williams College, Williamstown, MA 01267, USA\\
$^3$ Dipartimento di Scienze Fisiche, Universit\`{a} di Napoli
Federico II, Compl.
Univ. Monte S. Angelo, 80126 Naples, Italy  \\
$^4$ I.N.F.N., Sez. di Napoli, Complesso Universitario di Monte
Sant' Angelo, Edificio G, via Cinthia, 80126 - Napoli, Italy }
\date{Accepted xxx, Received yyy, in original form zzz}
\begin{document}
\maketitle
\begin{abstract}

The correlation between the peak photon energy of the internal spectrum $E_{\rm p,i}$ and  isotropic equivalent radiated energy $E_{\rm iso}$ (the \textit{Amati relation}) is explored in a scalar field  model of dark energy. Using an updated data set of $109$ high
redshift GRBs, we show that the correlation parameters only weakly
depend on the cosmological model. Once the parameters of Amati relation have been determined we use this relation to construct a \textit{fiducial}  GRBs Hubble diagram that extends up to redshifts $\sim 8$. Moreover we apply a local
regression technique to estimate, in a model independent way, the
distance modulus from the recently updated Union SNIa sample,
containing 557 SNIa spanning the redshift range of
$0.015 \le z \le 1.55$.
The derived calibration parameters are used to construct an updated GRBs Hubble diagram, which we
call the \textit{calibrated} GRBs HD. We also compare the
\textit{fiducial} and \textit{calibrated} GRBs HDs, which turned out to be  fully statistically consistent,
thus indicating that they are not affected by any systematic
bias induced by the different calibration procedures. This means that the high redshift GRBs can be used to
test different models of dark energy settling the circularity problem. Furthermore,  we investigate possible evolutionary effects that might have important influence on our
results.
Our analysis indicates that the presently available GRBs datasets do not show
\textit{statistically unambiguous} evolutionary effect with the cosmological redshift.
Finally we propose another approach to calibrate the GRB relations, by using an approximate
luminosity distance relation, which holds in any cosmological model.
%%depending only on the \textit{shape} of that function, which is a model
%independent manner to reconstruct the luminosity distance in a wide range of redshift.
We use this calibration of the Amati relation to construct an \textit{empirical approximate}  HD, which we compare with the \textit{calibrated} GRBs HD. We finally investigate the implications of this approach for the high redshift \textit{cosmography}.
\end{abstract}
\section{Introduction}
At the end  of the '90s observations of high redshift supernovae of type Ia (SNIa) revealed
that the universe is now expanding at an accelerated rate. This
surprising result has been independently confirmed by observations
of small scale temperature anisotropies of the cosmic microwave
background radiation (CMB) \cite{Riess07,SNLS,Union,WMAP3}. It is
usually assumed that the observed accelerated expansion is caused by
a so called dark energy, with unusual properties. The pressure of
dark energy $p_{de}$ is negative and it is related \textbf{to} the positive
energy density of dark energy $\epsilon_{de}$ by
$p_{de}=w\epsilon_{de}$ where the proportionality coefficient $w<0$.
According to the present day estimates, about 75\% of matter-energy
in the universe is in the form of dark energy, so that now the dark
energy is the dominating component in the universe. The nature of
dark energy is not known. Proposed so far models of dark energy can
be divided, at least, into three groups: a) a non zero cosmological
constant, in this case $w=-1$,  b) a potential energy of some not
yet discovered scalar field, or c) effects connected with non
homogeneous distribution of matter and averaging procedures. In the
last two possibilities, in general, $w$ is not constant and it
depends on the redshift $z$. Observations of type Ia supernovae and
small scale anisotropies of the cosmic microwave background
radiation are consistent with the assumption that the observed
accelerated expansion is due to the non zero cosmological constant.
However, so far the type Ia supernovae have been observed only at
redshifts $z<2$, while in order to test if $w$ is changing with
redshift it is necessary to use more distant objects. New
possibilities opened up when  the Gamma
Ray Bursts have been discovered at  higher redshifts, the present record is
at $z=8.26$ \cite{Greiner08}.
GRBs are however  enigmatic objects. First of all the mechanism
that is responsible for releasing the incredible amounts of energy
that a typical GRB emits is not yet known (see for instance Meszaros
2006 for a recent review). It is also not yet definitely
known if the energy is emitted isotropically or is beamed. Despite
of these difficulties GRBs are promising objects that can be used to
study the expansion rate of the universe at high redshifts
\cite{Bradley03,S03,Dai04,Bl03,Firmani05,S07,Li08,Amati08,Ts09}.
Using the observed spectrum and light curves it is possible to derive additional parameters,
for example the peak photon energy $E_{p,i}$, at which the burst is
the brightest, and  the variability parameter
$V$ which measures the smoothness of the light curve (for
definitions of these and other parameters mentioned below, see
Schaefer (2007)). From observations of the afterglow it is possible
to derive another set of parameters, redshift is the most important,
and also the jet opening angle $\Theta_{jet}$, inferred from the achromatic break  in the light curve of the afterglow, the time lag
$\tau_{lag}$, which measures the time offset between high and low
energy GRB photons arriving at the detector, and $\tau_{RT}$ - the
shortest time over which the light curve increases by half of the
peak flux of the burst. Moreover, even though the most important parameters - the
intrinsic luminosity $L$ and the total (isotropic) radiated energy $ E_{\rm iso}$ -
are not directly observable, several correlation relations have been found between the \textit{additional parameters}, like the peak photon energy $E_{p,i}$, the variability $V$, the time lag $\tau_{lag}$, etc.., and the GRBs radiated energy or luminosity. Assuming that GRBs emit radiation
isotropically it is possible to relate $L$  to the observed
bolometric peak flux $P_{bolo}$ and the bolometric fluence $S_{bolo}$ respectively by
\begin{equation}
L= 4\pi{d^2}_{L}(z, cp)P_{bolo},
\label{lum}
\end{equation}
and
\begin{equation}\label{eiso}
E_{\rm iso}=4\pi{d^2}_{L}(z, cp)S_{bolo}(1+z)^{-1},
\end{equation}
 where $d_{L}$ is the
luminosity distance, and $cp$ denotes the set of cosmological
parameters that specify the background cosmological model.
Equivalently one can use the total collimation corrected energy. It is clear
from these relations that, in order to get the intrinsic luminosity,
it is necessary to specify the fiducial cosmological model and its
basic parameters. But we want to use the observed properties of GRBs
to derive the cosmological parameters. Several procedures to overcome this
complicated circular situation have been proposed (see for instance \cite{S07,BP08,CCD,MEC10}).
In this paper we apply a Bayesian motivated technique, already implemented in \cite{MEC10} to standardize $109$ long GRBs  with the well-known
Amati relation, i.e. to find the $E_{\rm p,i}$ -- $E_{\rm iso}$ correlation parameters,
where $E_{\rm p,i}$  is the peak photon energy of the intrinsic spectrum and $E_{\rm iso}$ is the isotropic equivalent radiated energy, in order to construct an \textit{ estimated fiducial} Hubble diagram that extends to  redshifts $z\sim 8$, assuming that radiation propagates in a quintessential cosmological model. We begin our analysis using a minimally coupled self-interacting scalar field quintessense model, with an exponential
potential. Parameters of this model are fixed by fitting appropriate estimates to the set of
type Ia supernovae data, the power spectrum of CMB temperature
anisotropies and parameters of the observed large scale structure (see \cite{MECC}), which has been recently referred as \textit{ fiducial} model to  standardize the GRBs and to construct the Hubble diagram using previous samples of GRBs compiled by Schaefer 2007, Dainotti et al. 2008,  Amati et al. 2008 - 2009, and \cite{MEC10}. Here we are extending this analysis by considering new updated dataset.
Moreover we apply a local regression technique to estimate, in a model independent way, the
distance modulus from the recently updated SNIa sample,
referred to as Union \cite{Union2}, containing 557 SNIa spanning the redshift range
$0.015 \le z \le 1.55$. The derived calibration
parameters are used to construct a new  \textit{calibrated} GRBs Hubble diagram.
We also compare the \textit{estimated} and \textit{calibrated} GRBs HDs.  The scheme of the paper is as follows. In Section 2 we
describe our statistical method to fit the $E_{\rm p,i}$ -- $E_{\rm iso}$ correlation, to estimate the normalization and the slope of such a relation, and then construct the \textit{estimated} and \textit{calibrated} GRBs  HD. In Section 3  we  present an alternative procedure to calibrate the Amati relation in a cosmological-independent way, and we explore their implications for cosmography. Section 4 is devoted to discussion and conclusions.
\section{Standardizing the GRBs and constructing the Hubble diagram}
In this section we investigate the possibility of constructing the
Hubble diagram from the $E_{\rm p,i}$ -- $E_{\rm iso}$ correlation,
here  $E_{\rm p,i}$ is the peak photon energy of the intrinsic
spectrum and $E_{\rm iso}$ the isotropic equivalent radiated energy.
$E_{\rm iso}$ is defined by the Eq. (\ref{eiso}). This correlation was initially discovered in a small
sample of \sax{} GRBs with known redshifts \cite{Amati02} and
confirmed afterwards by HETE-2 and SWIFT observations \cite{Lamb05}, \cite{Amati06}. Although
it was the first correlation discovered for GRB observables it was never used
for cosmology because of its significant "extrinsic" scatter.
However, the recent increase in the efficiency of GRB  discoveries
combined with the fact that $E_{\rm p,i}$ -- $E_{\rm iso}$
correlation needs only two parameters that are directly inferred
from observations (this fact minimizes the effects of systematics
and increases the number of GRBs that can be used by a factor $\sim
3$) makes this correlation an interesting tool for
cosmology. Previous analyses of the \epeiso{} plane of GRBs parameters
showed that different classes of GRBs exhibit different behaviours,
and while normal long GRBs and X--Ray Flashes (XRF, i.e.
particularly soft bursts) follow the $E_{\rm p,i}$ -- $E_{\rm iso}$
correlation, short GRBs and the peculiar very near and
sub--energetic GRBs do not (Amati et al. 2008). This fact may depend  on
the different emission mechanisms involved in different classes of
GRBs and makes the \epeiso{} relation a useful tool to distinguish
between them \cite{Antonelli09}.  The impact of selection and instrumental
effects on the \epeiso{} correlation of long GRBs was investigated
since 2005, mainly based on the large sample of BATSE GRBs with
unknown redshifts. Different authors came to different conclusions
(see for instance \cite{Ghirlanda05}). In particular,
\cite{Ghirlanda05} showed that BATSE events potentially follow the
\epeiso{} correlation and that the question to clarify is if, and
how much, its measured dispersion is biased. There were also claims
that a significant fraction of \swift{} GRBs is inconsistent with
this correlation \cite{Butler07}. However,  when considering those
\swift{} events with peak energy measured by broad--band instruments
like, e.g., Konus--WIND or the \fermi/GBM  or reported by the BAT
team in their catalog \cite{Sakamoto08} it is found that they are
all consistent with the \epeiso{} correlation as determined with
previous/other instruments \cite{Amati09}. In addition, it turns out
that the slope and normalization of the correlation based on the  data
sets provided by GRB detectors with different sensitivities and energy
bands are very similar. These facts further support
the reliability of the Amati correlation \cite{Amati09}. It is clear
from Eq. (\ref{eiso}) that, in order to get $E_{\rm iso}$,
it is necessary to specify the fiducial cosmological model. In \cite{MEC10}, we fitted the Amati relation in a quintessence cosmological model, where the dark energy is described by the exponential potential of the scalar field discussed in \cite{MECC}.
In our analysis we consider a sample of 109 long GRB/XRF, adding to the  sample of 95 long GRB/XRF  compiled in (Amati et al. 2008) and (Amati et al. 2009) data of 14 unpublished GRBs, kindly provided by Amati in a private communication. Their redshift distribution covers a broad range of $z$, from $0.033$ to $8.23$, thus extending far beyond that of SNIa (z $<$$\sim$1.7), and including GRB $092304$, the new high-z record holder of Gamma-ray bursts.
\subsection{Fitting the $E_{\rm p,i}$ -- $E_{\rm iso}$ relation and estimating its parameters}
\begin{figure}
\includegraphics[width=8 cm]{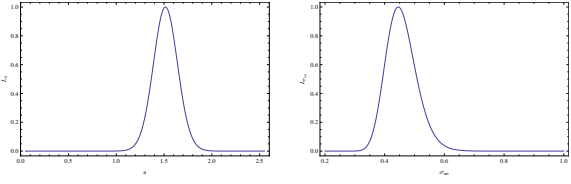}
\caption{Marginalized likelihood functions:  the likelihood function,$ {\cal{L}}_a$  is obtained marginalizing over $\sigma_{int}$; and the likelihood function,  ${\cal{L}}_{\sigma_{int}}$, is obtained marginalizing over $a$.}
\label{likelihoods}
\end{figure}

\begin{figure}
\includegraphics[width=7 cm]{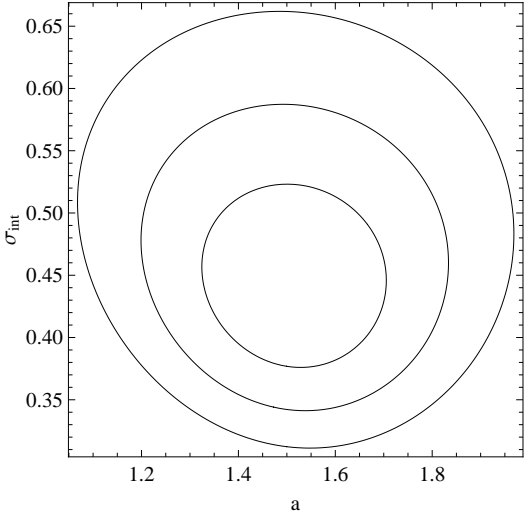}
\caption{Regions of  $68\%$, $95\%$ and $99\%$ of confidence in the
space of parameters $a,\sigma_{int}$.}
\label{confidencereg_egepisoscal}
\end{figure}

\begin{figure}
\includegraphics[width=7 cm]{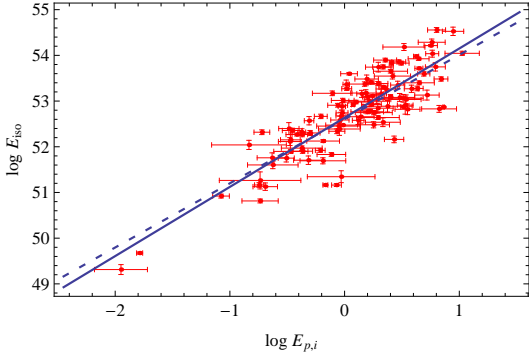}
\caption{Best fit curves for the  $E_{\rm p,i}$ -- $E_{\rm iso}$ correlation relation superimposed on
the data. The solid and dashed lines refer to the results
obtained with the maximum likelihood and weighted $\chi^2$ estimator
respectively.}
\label{eg-episocorr_scal}
\end{figure}

\begin{figure}
\includegraphics[width=7 cm]{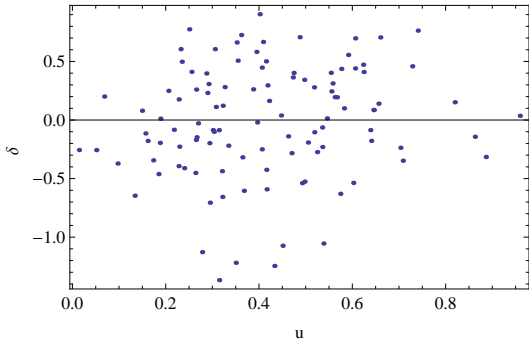}
\caption{Behaviour with the redshift (\textbf{in a logarithmic scale, being $u=\log(1+z)$}) of the residuals, $\delta =y_{obs} - y_{fit}$, for the $E_{\rm p,i}$ -- $E_{\rm iso}$  correlation relation. We see that the figure does not exhibit any evidence of correlation. }
\label{residuals}
\end{figure}

In this section we present results of our
analysis of the Amati correlation performed on a new updated dataset, assuming that the
background cosmological model is one of the quintessence models that we have studied some time ago \cite{MECC}, and showed that it is  consistent with the basic cosmological tests, and which has been recently referred as \textit{ fiducial} model to  standardize the GRBs and to construct the Hubble diagram using previous samples of GRBs.
First of all, we consider the $E_{\rm p,i}$ -- $E_{\rm iso}$ relation in the form
 \begin{eqnarray}
 \label{eqamati}
  \log \left(\frac{E_{\rm iso}}{1\;\mathrm{erg}}\right)  &=& b+a \log  \left[
    \frac{E_{\mathrm{p, i}} (1+z)}{300\;\mathrm{keV}}
  \right]\,,
\end{eqnarray}
where $a$ and $b$ are constants.
In fitting this relation, we need to fit a data
array $\{x_i, y_i\}$ with uncertainties $\{\sigma_{x,i},
\sigma_{y,i}\}$, to a straight line
\begin{equation}
\label{linear} y =b + a x\,,
\end{equation}
in order to determine the two fit parameters $(a, b)$. Actually, the situation is not so
simple since, both the $(y, x)$ variables are affected by
measurement uncertainties $(\sigma_x, \sigma_y)$ which can not be
neglected. Moreover, $\sigma_y/y \sim \sigma_x/x$ so that it is
impossible to choose as independent variable in the fit the one
with the smallest relative error. Finally, the  correlation we are
fitting is not of theoretical nature, i.e, it is not (yet) derived from an
underlying theoretical model determining the detailed features of
the GRBs explosion and afterglow phenomenology. Indeed, we do expect
a certain amount of intrinsic scatter, $\sigma_{int}$, around the best fit line that
has to be taken into account and determined together with $(a, b)$ by
the fitting procedure. Different statistical recipes are available
to cope with these problems.
As in \cite{MEC10}, we apply a Bayesian motivated
technique \cite{dagostini}  maximizing the likelihood function
${\cal{L}}(a, b, \sigma_{int}) = \exp{[-L(a, b, \sigma_{int})]}$
with\,:
\begin{eqnarray}
L(a, b, \sigma_{int}) & = & \frac{1}{2} \sum{\ln{(\sigma_{int}^2 +
\sigma_{y_i}^2 + a^2
\sigma_{x_i}^2)}} \nonumber \\
~ & + & \frac{1}{2} \sum{\frac{(y_i - a x_i - b)^2}{\sigma_{int}^2 +
\sigma_{x_i}^2 + a^2 \sigma_{x_i}^2}}\,, \label{eq: deflike}
\end{eqnarray}
where the sum is over the ${\cal{N}}$ objects in the sample. Note
that, actually, this maximization is performed in the two parameter
space $(a, \sigma_{int})$ since $b$ may be estimated analytically by solving the equation $\displaystyle {\frac{\partial }{\partial b}L(a, b, \sigma_{int})=0}$, \,as\,:
\begin{equation}
b = \left [ \sum{\frac{y_i - a x_i}{\sigma_{int}^2 + \sigma_{y_i}^2
+ a^2 \sigma_{x_i}^2}} \right ] \left [\sum{\frac{1}{\sigma_{int}^2 + \sigma_{{y_i}}^2 + a^2 \sigma_{x_i}^2}} \right ]^{-1}\,. \label{eq:calca}
\end{equation}
To quantitatively estimate the goodness of this fit we use
the median and root mean square of the best fit residuals, defined
as $\delta = y_{obs} - y_{fit}$.
To quantify the uncertainties of some fit parameter $p_i$, we
evaluate the marginalized likelihood ${\cal{L}}_i(p_i)$  by
integrating over the other parameter. The median value for the
parameter $p_i$ is then found by solving\,:
\begin{equation}
\int_{p_{i,min}}^{p_{i,med}}{{\cal{L}}_i(p_i) dp_i} = \frac{1}{2}
\int_{p_{i,min}}^{p_{i,max}}{{\cal{L}}_i(p_i) dp_i} \ . \label{eq:
defmaxlike}
\end{equation}
The $68\%$ ($95\%$) confidence range $(p_{i,l}, p_{i,h})$ are then
found by solving \cite{dagostini}\,:
\begin{equation}
\int_{p_{i,l}}^{p_{i,med}}{{\cal{L}}_i(p_i) dp_i} = \frac{1 -
\varepsilon}{2} \int_{p_{i,min}}^{p_{i,max}}{{\cal{L}}_i(p_i) dp_i}
\ , \label{eq: defpil}
\end{equation}
\begin{equation}
\int_{p_{i,med}}^{p_{i,h}}{{\cal{L}}_i(p_i) dp_i} = \frac{1 -
\varepsilon}{2} \int_{p_{i,min}}^{p_{i,max}}{{\cal{L}}_i(p_i) dp_i}
\ , \label{eq: defpih}
\end{equation}
with $\varepsilon = 0.68$ and $\varepsilon = 0.95$ for the $68\%$
and $95\%$ confidence level. Just considering  our correlation in Eq. ( \ref{eqamati})
we find that the likelihood method gives
$a=1.52$, $b=52.67$ and $\sigma_{int}=0.41$.
In Fig. \ref{confidencereg_egepisoscal} we show
the likelihood contours in the $(a, \sigma_{int})$ plane and
in Fig. \ref{eg-episocorr_scal} we show the correlation between the
observed $\log{E_{\rm p,i}}$ and derived $\log{E_{\rm iso}}$  with our
assumed background cosmological model. The solid line is the best
fit obtained using the D'Agostini's method \cite{dagostini} and the
dashed line is the best fit obtained by the weighted $\chi^2$
method. If one \textit{marginalizes}\footnote{It is worth noting that in the marginalization procedure we have to take into account also the Eq. \ref{eq:calca}.} with respect to $b$, then the likelihood
values of $a$ and $\sigma_{int}$ are $a=1.52^{+0.13}_{-0.19}$ and $\sigma_{int}=0.45\pm 0.05$. The marginalized likelihood functions are shown in Fig. \ref{likelihoods}.
The performed statistical analysis shows that the relation (\ref{eqamati}) has a statistical weight similar to the one exhibited by the other relations previously studied in \cite{MEC10}, since both
$\delta_{med}$  and $\delta_{rms}$ have almost the same values over
the full set (of relations), where $\delta =y_{obs} - y_{fit}$.
In our analysis we have assumed that the fit parameters do not change with the
redshift, which indeed spans a quite large range (from $z = 0.0331$
up to $z \simeq 8$). The limited number of GRBs prevents detailed
exploration of the validity of this usually adopted working
hypothesis, which we tested somewhat investigating  if the residuals
correlate with the reshift. We have not found any significant
correlation, as shown in Fig. \ref{residuals}. Moreover we tested the fit of the
$E_{\rm p,i}$ -- $E_{\rm iso}$  correlation relation with respect to the evolution with redshift,
separating the GRB samples  into four groups corresponding to the
following redshift  bins: $z\in [0,1]$, $z\in (1,2]$, $z\in (2,3]$
and  $z \,>\, 3$. We thus maximized the likelihood in each group of
redshifts and determined the best fit parameters $a$, $b$
at $65\%$  confidence level, and the intrinsic dispersion $\sigma_{int}$, as summarized in Table 1.
It turns out that no statistical evidence of a dependence of the $(a, b,\sigma_{int})$ parameters on the redshift exists. This is in agreement with what has recently been found by Ghirlanda et al. (2008) and Wang, Deng and Qiu (2008), or, as regards to other correlation relations, by Cardone et al. (2008) and Basilakos \& Perivolaropoulos (2008), and also confirmed in our previous paper \cite{MEC10}.
\begin{table}
\caption{Calibration parameters $(a, b)$, intrinsic scatter $\sigma_{int}$, median, $\delta_{med}$, and root mean square of the best fit residuals, $\delta_{rms}$ for the Amati correlation, evaluated in four ranges of redshift  ($z\in (0,1]$, $z\in (1,2]$, $z\in (2,3]$
and $z \,>\, 3$, where however we have only 10 GRBs).
Columns are as follows\,: 1. id of the reshift range;  2. maximum likelihood parameters;  $3$\,,$4$ $68\,\%$ confidence ranges for the parameters $(a, \sigma_{int})$;  $5$\,,$6$ median and root mean square of the residuals. }
\begin{center}
\begin{tabular}{|c|c|c|c|c|c|}
\hline Id & $(a, b, \sigma_{int})_{ML}$  &
$(a_{-1\sigma }, a_{+1\sigma })$ & $((\sigma_{int})_{-1\sigma
}, \sigma_{{int}_{+1\sigma}})$ & $\delta_{med}$ &
$\delta_{rms}$ \\ \hline \hline
 ~ & ~ & ~ & ~&  ~ & \\
$z\in [0,1]$ &  (1.53, 52.6, 0.36) &
$(1.28\,, 1.69)$ & $(0.27\,,0.51)$ & $-0.07 $&$0.39$\\
 ~ & ~ & ~ & ~&  ~ & \\
$z\in (1,2]$ &  (1.38, 52.6, 0.55)&
$(0.8, 1.9)$ & $(0.43\,,0.7)$ &$0.08$&$0.58$\\
 ~ & ~ & ~ & ~&  ~ & \\
$z\in (2,3]$ &  (1.38, 52.64, 0.56) &
$(0.83\,,1.94)$ & $(0.43,0.75)$ &$0.08$&$0.59$\\
 ~ & ~ & ~ & ~&  ~ & \\
$z>3$ & (1.59, 52.9, 0.22)&
$(1.08\,, 2.01)$ & $(0.1\,,0.46)$ &$0.01$&$0.29$\\
~ & ~ & ~ & ~&  ~ & \\
\hline
\end{tabular}
\end{center}
\end{table}
\subsection{Constructing the Hubble diagram}
\begin{figure}
\includegraphics[width=7 cm]{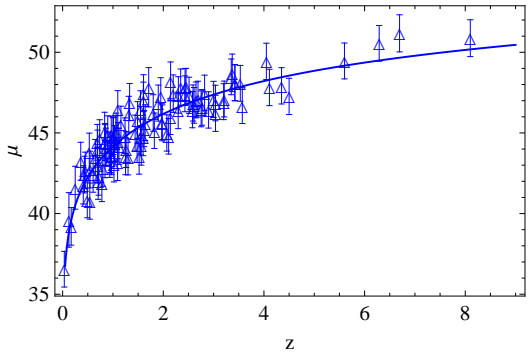}
\caption{The  \textit{estimated} quintessential Hubble diagram, with superimposed (solid line) the theoretical curve.}
\label{hdgrbamati}
\end{figure}

Once the Amati correlation relation has been fitted, and its parameters have been estimated, we can now use them
to construct the \textit{estimated fiducial} GRBs Hubble diagram. Actually let us remind that the luminosity distance
of a GRB with the redshift $z$ may be computed as\,:
\begin{equation}\label{lumdist}
d_L(z)^2 = \left( \frac{E_{\rm iso}(1 + z)}{4 \pi  S_{bolo}}\right).
\end{equation}
The uncertainty of $d_L(z)$ is then estimated through the propagation of the measurement errors on the involved
quantities. In particular, remembering that our correlation relation  can be written as a linear relation, as in Eq. (\ref{linear}), where $y$ is  the distance dependent quantity, while $x$ is not, the error on the distance
dependent quantity $y$ is estimated as:
\begin{equation}
\sigma(\log{y}) = \sqrt{a^2 \sigma^2(\log{x}) + \sigma_{int}^2}\,,
\label{eq: siglogy}
\end{equation}
and is then added in quadrature to the uncertainties of the other terms entering
Eq.(\ref{lumdist}) to get the total uncertainty. The distance modulus $\mu(z)$ is
easily obtained from its definition\,:
\begin{equation}
\mu(z) = 25 + 5 \log{d_L(z)}\,, \label{eq:defmu}
\end{equation}
with its uncertainty obtained again by error propagation. We finally estimate the
distance modulus for each $i$\,-\,th GRB in our sample at redshift
$z_i$, to build the Hubble diagram plotted in Fig.
\ref{hdgrbamati}. In what follows we  will refer to this data set as the
{\it fiducial} GRBs Hubble diagram (hereafter, HD) since to compute the
distances it relies on the calibration based on the fiducial quintessential model.
We also investigated the impact of varying the parameters of our fiducial cosmological model, fitting the Amati
correlation relation on a regular grid in the space of parameters of our quintessential model, as allowed by the data \cite{MECC}.
For each point in the grid, we repeated all the steps described above to get the distance modulus to each GRB in the sample. We then collect these values and evaluate, for each GRB, the root mean square of the percentage deviation from the fiducial $\mu$ value.  In this way we are performing a sort of average of the absolute percentage deviation, which allows us to quantify the order of magnitude of the studied effect. Actually it turns out that the distance modulus may be under or overestimated by a modest $0.3\%$ with values never larger than $1\%$.
\subsubsection{Cosmological independent calibration: local regression on SNIa}
\begin{figure}
\includegraphics[width=7 cm]{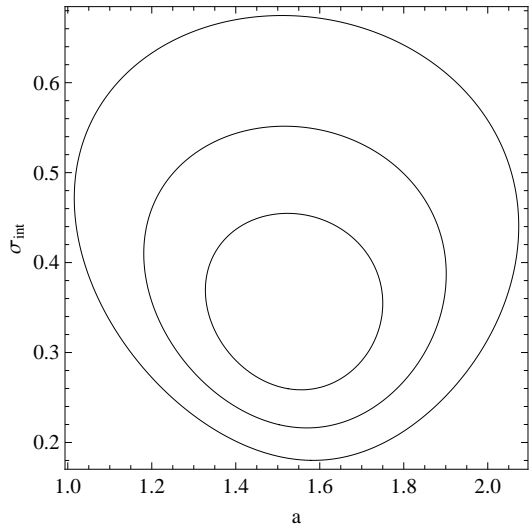}
\caption{Regions of  $68\%$, $95\%$ and $99\%$ of confidence in the
space of parameters $a,\sigma_{int}$, obtained by the local regression on SNeIa.}
\label{confidencereg_egepisoscalSNIA}
\end{figure}

\begin{figure}
\includegraphics[width=8 cm]{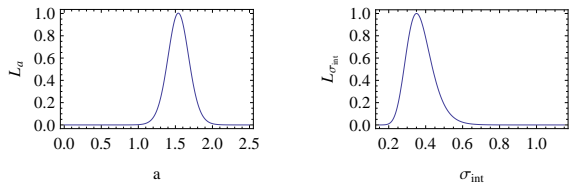}
\caption{Marginalized likelihood functions constructed by the local regression on SNeIa:  the likelihood function,$ {\cal{L}}_a$  is obtained marginalizing over $\sigma_{int}$; and the likelihood function,  ${\cal{L}}_{\sigma_{int}}$, is obtained marginalizing over $a$.}
\label{likelihoodsnia}
\end{figure}

Although the above analysis has shown that the choice of the
underlying cosmological model has only a modest impact on the final
estimate of the distance modulus, we compared our \textit{estimated fiducial HD} with a
model independent calibrated HD, carried out using SNIa as distance
indicators. We apply local regression to estimate the distance modulus $\mu(z)$ from
the recently updated SNIa sample, the SCP Union2 compilation \cite{Union2}, which is an update of the original Union compilation, now bringing together data for 719 SNe, drawn from $17$ datasets. Of these, 557 SNe, spanning the range $0.015 \le z \le 1.55$, form the final sample considered in our analysis. To use this large dataset as input to the local regression estimate of $\mu(z)$ we have firstly to set a redshift $z_i$ where $\mu(z_i)$ has to be recovered and we order the SNIa dataset  according to increasing value of $|z - z_i|$ and we select the first $n = \alpha {\cal{N}}_{SNIa}$, where $\alpha$ is a user selected value and ${\cal{N}}_{SNIa}$ the total number of SNIa. Then we can fit a first order polynomial to the previously selected data, weighting each SNIa with the corresponding value of an appropriate weight function, like, for instance
\begin{equation}
W(u) = \left \{
\begin{array}{ll}
(1 - |u|^2)^2 & |u| \le 1 \\ ~ & ~ \\ 0 & |u| \ge 1\,,
\end{array}
\right .
\label{eq: wdef}
\end{equation}
and take the
zeroth order term as the best estimate of $\mu(z)$. Here  $u = |z - z_i|/\Delta$ and $\Delta$ is the maximum value of the $|z -
z_i|$ over the subset chosen before. To estimate the error on $\mu(z)$ we use the root mean square of the weighted
residuals with respect to the best fit zeroth order term \footnote{
It is worth stressing that both the choice of the weight function and the order
of the fitting polynomial are somewhat arbitrary. Similarly, the value of $\alpha$
to be used must not be too small in order to make up a statistically valuable sample,
but also not too large to prevent the use of a low order polynomial. In our local regression routine we have performed
an extensive set of simulations, ending up with a mock catalogue having the same redshift and error distribution of the actual SNIa one. This mock
catalogue is used as input to the routine sketched above and finally the reconstructed
$\mu(z)$ value for each point in the catalog are compared to the input one (see \cite{CCD} for details).}.
Having found an efficient way of estimating the distance modulus
at redshift $z$ in a model independent way, we can now fit
the $E_{\rm p,i}$ -- $E_{\rm iso}$  correlation relation, using the \textit{local regression reconstructed}
$\mu(z)$ in Eq. (\ref{eiso}).
We consider only GRBs with $z \le 1.55$ in order to cover the same redshift range spanned by the
SNIa data. For such subset of GRBs we apply the Bayesian fitting
procedure described above to estimate the correlation parameters. We use the other GRBs to construct a new GRBs Hubble diagram
that we call the {\it calibrated} GRBs HD. In Fig. \ref{confidencereg_egepisoscalSNIA} we show
the likelihood contours in the $(a, \sigma_{int})$ plane, and the marginalized likelihood functions are shown in Fig. \ref{likelihoodsnia}.

\begin{figure}
\includegraphics[width=6 cm]{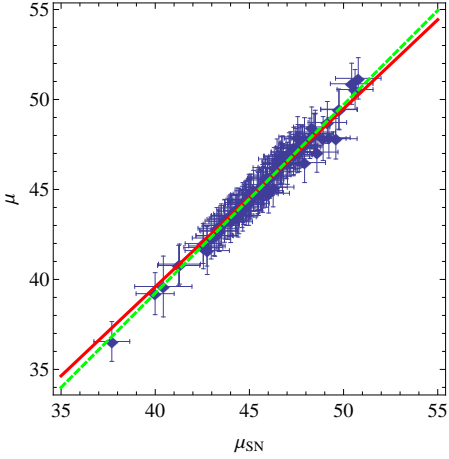}
\caption{Comparison of the distance modulus $\mu(z)$ for the
\textit{calibrated} and \textit{estimated} GRBs Hubble diagram made up fitting the Amati correlation. The red solid line represent the graph of the function $F(\mu_{SN})=\mu_{SN}$, and it turns out that it is also the fitting function if we do not include a constant term in the list of basis functions (nedeed to perform the fit procedure). The dashed green line is the best fit function.}
\label{hdcalfidamatnew2}
\end{figure}

 Moreover it turns out that the \textit{estimated} and \textit{calibrated} HDs are fully statistically consistent, as shown in  Fig.\,\ref{hdcalfidamatnew2}.

\section{An alternative procedure to calibrate the $E_{\rm p,i}$ -- $E_{\rm iso}$ relation in a cosmological-independent way: implications for cosmography }
In this section we propose a procedure to calibrate the $E_{\rm p,i}$ -- $E_{\rm iso}$ relation using only the GRBs events with
the redshift $z \le 1.55$ without specifying a cosmological model.
The luminosity distance estimations to GRBs are inferred  from many known Type Ia supernovae, and based on an approximate formula for the luminosity distance which holds in any cosmological model, depending only on the \textit{shape} of this function, more than on a power series expansion in the redshift parameter z (the coefficients of such an expansion being functions of the scale factor a(t) and its higher order derivatives), as in the cosmographic approach. Our starting point is the relation between the angular diameter distance $D_A$ and the luminosity distance $D_L$
\begin{equation}\label{lum2}
D_{L}=\left(1+z\right)^2D_A,
\end{equation}
where the angular diameter distance $D_A$ is a solution of the equation
\begin{equation}   \left ( \frac{dz}{dv}
\right )^2 \frac{d^2 D}{dz^2} + \left (   \frac{d^2z}{dv^2} \right )
\frac{dD}{dz} +\frac{4\pi G}{c^{4}} T_{\alpha   \beta} k^{\alpha} k^{\beta} D = 0.
\label{eq:angdiam2}
\end{equation}
with the following initial conditions:
\begin{eqnarray}\label{eq:cauchy}   &&D(z)\arrowvert_{z = 0} = 0,
\nonumber\\ && \\ &&\frac{dD(z)}{dz}   \arrowvert_{z = 0} =
\frac{c}{H_0}.\nonumber
\end{eqnarray}
In Eq. (\ref{eq:angdiam2}) $\it{v}$ is the affine parameter, $T_{\alpha   \beta}$ is the matter density tensor,  $k^{\alpha}=\displaystyle{{dx^{\alpha}\over dv}}=-\Sigma,_{\alpha}$ is the vector field tangent to the light ray congruence, and $\Sigma$ is the null surface along which the light rays propagate from the source.
In the general form the Eq.~(\ref{eq:angdiam2}) is very
complicated. General properties of this equation have been extensively studied (see for instance, \cite{Kant98,Kant20,Kant01,approx}). In most cases equation (\ref{eq:angdiam2}) does not have analytical solution, and from the mathematical point of view it can be reduced to a Fuchsian type with several regular singular points and a regular singular point at infinity. The solutions near each of these singular points, can be expanded in a series of hypergeometric functions. When we introduced the dimensionless angular diameter distance
 $r=DH_{0}/ c$ we discovered \cite{approx} that there is a simple function $r(z)$, which quite accurately
reproduces the exact numerical solutions of the equation (\ref{eq:angdiam2})  for $z$ up to very high values, it has the form
\begin{equation}
r(z)=\frac{z}{{\sqrt{d_1\,z^2 +
       {\left( 1 + d_2\,z +
          d_3\,z^2 \right) }^2}}},
\label{eq:approx}
\end{equation}
where $d_1$, $d_2$ and $d_3$ are constants.
Moreover the function (\ref{eq:approx}) automatically
satisfies the imposed initial conditions, so $r(0)=0$ and
$\displaystyle{dr \over dz}(0)=1$.
This approximate expression immediately provides an empirical formula for the luminosity distance relation
of the type Ia supernovae, through the Eq. (\ref{lum}). Fitting the relation (\ref{eq:approx}) to the SNIa Union dataset, we obtained the following best fit values for the fitting parameters  $d_i:$

\begin{eqnarray}
d_1 &=& -3.87\label{d1d2d3}\\
 d_2 &=& 1.51\label{d1d2d3b}\\
d_3 & =& 0.70.\label{d1d2d3c}
\end{eqnarray}

\begin{figure}
\includegraphics[width=7 cm]{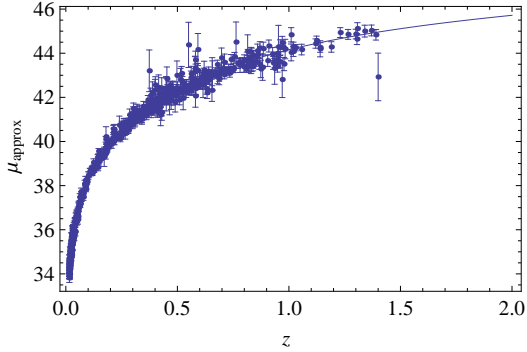}
\caption{Behaviour of the approximate modulus of distance with the redshift, with superimposed the SNIa Union dataset.   }
\label{approxihdsnia}
\end{figure}

\begin{figure}
\includegraphics[width=7 cm]{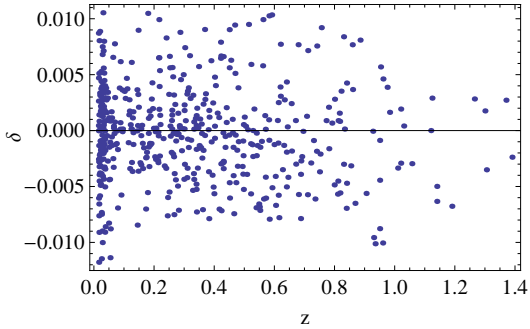}
\caption{Behaviour with the redshift of the \textit{relative residuals}, $\delta =\frac{r_{obs} - r_{fit}}{r_{obs}}$, of the fit procedure which provides our empirical formula in Eq. (\ref{eq:approx}), which agrees with the real data within a relative error of magnitude not more than few $~\%$. }
\label{residuals2}
\end{figure}

In Fig.(\ref{approxihdsnia})  we show the approximate distance modulus with superimposed the SNIa data. The approximate function agrees with the real data within a relative error not larger than a few $~\%$ in the SNIa redshift range, as shown in Fig.(\ref{residuals2}).

\begin{figure}
\includegraphics[width=7 cm]{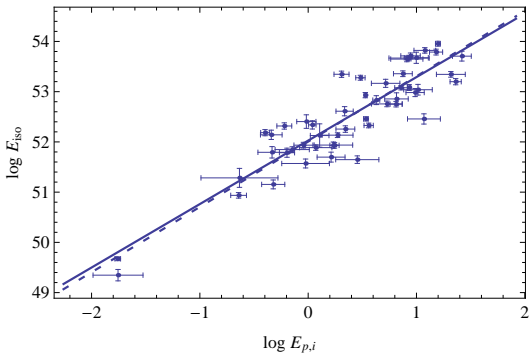}
\caption{Best fit curves for the  $E_{\rm p,i}$ -- $E_{\rm iso}$ correlation relation, obtained using our empirical approximate luminosity distance formula, superimposed on the data. The solid and dashed lines refer to the results
obtained with the Bayesian and Levenberg\,-\,Marquardt estimator
respectively.}
\label{eg-episocorr_approx}
\end{figure}

\begin{figure}
\includegraphics[width=7 cm]{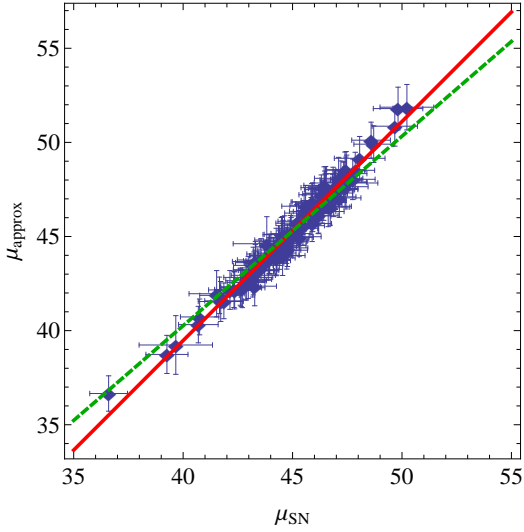}
\caption{Comparison of the distance modulus $\mu(z)$ for the
 \textit{empirical approximate} and \textit{calibrated} (SnIa) GRBs Hubble diagram. The dataset are fully statistically consistent are strongly correlated with the Spearman's correlation $\rho = 0.99$. The red solid line represent the graph of the function $F(\mu_{SN})=\mu_{SN}$, and it turns out that it is also the fitting function if we do not include a constant term in the list of basis functions (nedeed to perform the fit procedure). The dashed green line is the best fit function.}
\label{muappcal}
\end{figure}

\begin{figure}
\includegraphics[width=7 cm]{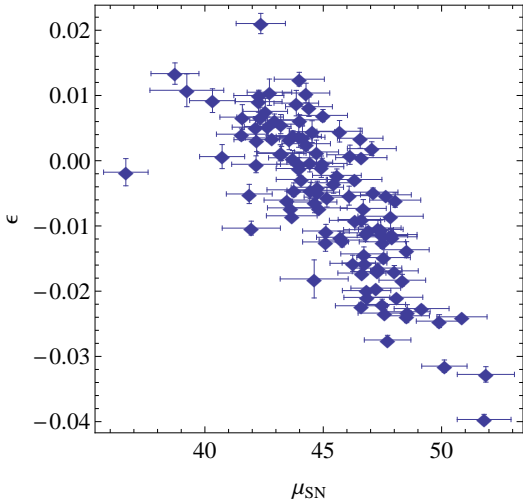}
\caption{Residuals (relative) $\epsilon= \frac{\mu_{SN} - \mu_{approx}}{\mu_{SN}}$ between the approximate and calibrated GRBs HDs.  }
\label{res_muappcal}
\end{figure}

From the \textit{empirical approximate} luminosity distance we can construct the  \textit{empirical approximate} distance modulus $\mu_{approx}(z)$, which we use to calibrate in a cosmological independent way, alternative to the one described above, the $E_{\rm p,i}$ -- $E_{\rm iso}$ (Amati) correlation relation, following our standard procedure, and considering only the GRBs with $z \le 1.55$ in order to cover the same redshift range spanned by the SNIa data. We use the other GRBs to construct a new GRBs Hubble diagram that we call the {\it approximate calibrated} GRBs HD.
In Fig. \ref{eg-episocorr_approx} we show the correlation between the
observed $\log{E_{\rm p,i}}$ and derived $\log{E_{\rm iso}}$  with our
approximate luminosity distance.
The solid line is the best
fit obtained using the D'Agostini's method \cite{dagostini} and the
dashed line is the best fit obtained by the weighted $\chi^2$
method. In order to check the reliability of the {\it approximate calibrated} GRBs HD we compare it with the {\it calibrated} (SNIa) HD.
It turns out that these HDs are fully statistically consistent, as shown in  Fig.\,\ref{muappcal} and \ref{res_muappcal}, and the resulting
distances are strongly correlated with the Spearman's correlation $\rho = 0.99$.
\subsection{Implications for high redshift cosmography}
Recently the cosmographic approach to cosmology  gained increasing interest for catching as much information
as possible directly from observations, retaining the minimal priors
of isotropy and homogeneity and leaving aside any other
assumptions. Actually, the only ingredient taken into
account \textit{a priori} in this approach  is the FLRW line element obtained from
kinematical requirements
\begin{equation}
ds^2=-c^2dt^2+a^2(t)\left[\frac{dr^2}{1-kr^2}+r^2d\Omega^2\right].
\end{equation}

\begin{figure}
\includegraphics[width=7 cm]{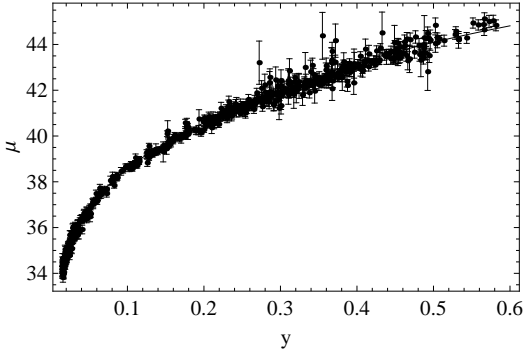}
\caption{Distance moduli for the best-fit values of our cosmography, together with the
  Union dataset. }
\label{cosmosneIa}
\end{figure}

\begin{figure}
\includegraphics[width=7 cm]{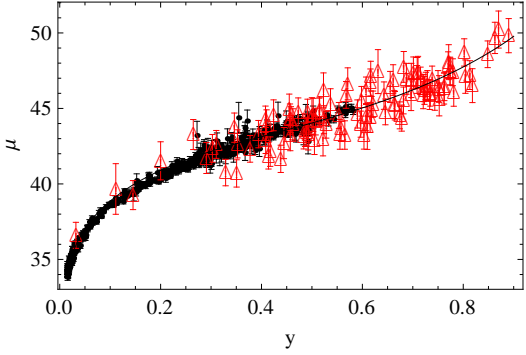}
\caption{Distance modulus for the best-fit values of our cosmography performed with GRBs only, together with the SneIa Union dataset (filled black circles)  and the \textit{calibrated} GBRs HD (empty red triangles). }
\label{cosmogrb}
\end{figure}

\begin{figure}
\includegraphics[width=7 cm]{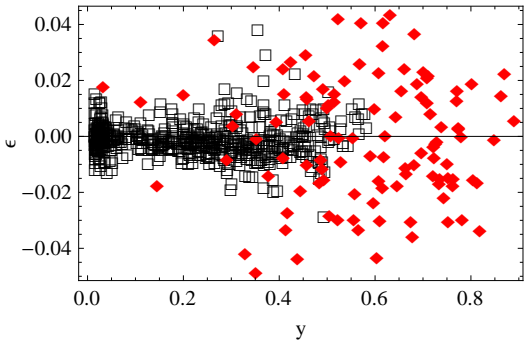}
\caption{Residuals (relative) $\epsilon= \frac{\mu_{obs} - \mu_{fit}}{\mu_{obs}}$ for our \textit{approximate cosmography}: the empty black boxes correspond to the SneIa Union dataset and the filled red diamond to the \textit{calibrated} GBRs HD.  }
\label{res_cosmogrb}
\end{figure}

Using this metric, it is possible to express the luminosity
distance $d_L$ as a power series in the redshift parameter $z$, the
coefficients of the expansion being functions of the scale factor
$a(t)$ and its higher order derivatives. Such an
expansions leads to a distance\,-\,redshift relation which only
relies on the assumption of the Robertson\,-\,Walker metric thus
being fully model independent since it does not depend on the
particular form of the solution of cosmic evolution equations. To this aim,
it is convenient to introduce the following parameters \cite{Visser}:
\begin{eqnarray}
H &=& \frac{1}{a} \frac{da}{dt}\,, \\
q &=& - \frac{1}{a} \frac{\frac{d^2a}{dt^2} }{ H^{2}}\,,\\
j &=& \frac{1}{a} \frac{\frac{d^3a}{dt^3} }{ H^{3}}\,, \\
s &=& \frac{1}{a} \frac{\frac{d^4a}{dt^4} }{ H^{4}}\,, \\
\label{eq: cosmopar}
\end{eqnarray}
which are usually referred to as the {\it Hubble, deceleration, jerk}, and {\it snap}
parameters, respectively\footnote{Note that the
use of the jerk parameter to discriminate between different models was also
proposed in \cite{SF} in the context of the {\it statefinder}
parametrization.}. Their present day values (which we will denote with a
subscript $0$) may be used to characterize the evolutionary status of the
Universe. For instance, $q_0 < 0$ denotes an accelerated expansion, while
$j_0$ allows to discriminate among different accelerating models.
It is worth noting that it is possible to infer implications for cosmography just using our empirical formula (\ref{eq:approx}) of the luminosity distance in our analysis. Actually we first recast our approximate  $d_L$ as a function of a
new variable $y=z/(1+z)$~ \cite{vitagliano,izzo} in such a way that $z\in(0,\infty)$ is mapped into
$y\in(0,1)$, obtaining
\begin{equation}\label{approxl-y}
d_L^{approx}(y)=\frac{c}{H_0}\left(-\frac{y}{(y-1)^3 \sqrt{\frac{d_1
   y^2}{(y-1)^2}+\left(-\frac{d_2 y}{y-1}+\frac{d_3
   y^2}{(y-1)^2}+1\right)^2}}\right).
   \end{equation}
Expanding our approximate luminosity distance up to the fourth order
in the $y$-parameter, we get

\begin{equation}\label{approx-cosmo}
d_L(y)=\frac{c}{H_0}\left\{y^3 \left(-\frac{d_1}{2}+d_2^2-4
   d_2-d_3+6\right)+y^4 \left(\frac{1}{2} d_1 (3
d_2-5)- d_2^3+5 d_2^2+2 d_2(d_3-5)-5
   (d_3-2)\right)-(d_2-3) y^2+y\right\}~.
\end{equation}
It is now possible to relate the $d_i$ to the cosmographic parameters $q_0$, $j_0$, $s_0$, by comparing the expansion in Eq. (\ref{approx-cosmo}) with the standard expansion to the fourth order:
\begin{eqnarray}
d_L(y)=&&\frac{c}{H_0}\left\{y-\frac{1}{2}(q_0-3)y^2+\frac{1}{6}\left[12-5q_0+3q^2_0-j_0\right]y^3+\frac{1}{24}\left[60-7j_0-\right.\right.\nonumber
  \\ &&\left.\left.-10-32q_0+10q_0j_0+6q_0+21q^2_0-15q^3_0+s_0\right]y^4+\mathcal{O}(y^5)\right\}~.
\end{eqnarray}
It turns out that
\begin{eqnarray}
  d_1 &=& \frac{1 + j_0 + q_0 + 6 j_0 q_0 - 2 q_0^2 (1 + 3 q_0) + s_0}{6 (3 + q_0)}\label{d1eq} \\
  d_2 &=& \frac{3 + q_0}{2}\label{d2eq} \\
  d_3 &=& \frac{14 + j_0 (5 - 4 q_0) + q_0 (16 + 3 (-1 + q_0) q_0) - s_0}{12 (3 + q_0)}\label{d3eq}.
\end{eqnarray}
Inverting the systems of Eqs. (\ref{d1eq}, \ref{d2eq}, \ref{d3eq}) it is possible to recover the cosmographic parameters $q_0$, $j_0$ and $s_0$ as functions of our fitted parameters $d_i$. Actually we get
\begin{eqnarray}
  q_0 &=& -3+ 2 d_2\label{q0eq} \\
  j_0 &=& 17 + 3 d_1 - 22 d_2 + 6 d_2^2 + 6 d_3\label{j0eq} \\
  s_0 &=& d_1 (51 - 24 d_2) + 158 d_2^2 - 24 d_2^3 - 8 d_2 (35 + 9 d_3) +
 3 (49 + 34 d_3)\label{s0eq}.
\end{eqnarray}
These equations, together with the values of the fitting parameters  in Eqs. (\ref{d1d2d3}, \ref{d1d2d3b}, \ref{d1d2d3c}) with the corresponding \textit{physically acceptable} regions of confidence\footnote{It is worth noting that such \textit{physically acceptable} regions are obtained imposing, on the confidence regions provided by the standard statistical fitting procedure, some \textit{priors}, which have to guarantee that the approximate function in Eq. (\ref{eq:approx}) preserves the special shape typical of the angular diameter distance. Instead, since the SNIa dataset used to fit the parameters $d_i$ is limited in redshift by $z\leq 1.39$, the approximate function in Eq.(\ref{eq:approx}) is not sampled at higher values of the redshift, and its  behaviour  could result misshaped.  } allow us to estimate the corresponding parameter confidence intervals for $q_0$, $j_0$ and $s_0$. We actually get that
$\left(
\begin{array}{ccc}
q_0 & -1.11 & 0.27 \\
 j_0 & -23. 9& 1.70 \\
s_0 &  -256.4 & -34.7
\end{array}\label{q0j0s0}
\right)  $,
the best fit being $q_0=-0.45$, $j_0=-12.3$ and $s_0=-99.3$, which agree with the values found in literature (see for instance \cite{vitagliano,izzo}).
In Fig. (\ref{cosmosneIa}) we plot our \textit{approximate cosmographic distance moduli} together with the SNIa Union dataset.
In addition, we investigate the possibility to use high redshift GRBs to determine parameters of our  \textit{approximate cosmography}. Therefore, in the following we use the \textit{calibrated} (with SNIa) GBRs HD, described above, to fit the $d_i$ parameters and then to derive the cosmographic parameters $q_0$, $j_0$, and $s_0$.
We obtain the following ($2\sigma$) parameter confidence  intervals
$\left(
\begin{array}{ccc}
 d_1& 0.950028 & 11.2294 \\
 d_2&-2.76299 & 1.92514 \\
 d_3& 0.140305 & 0.836675\,
\end{array}
\right)$,
and it turns out that the corresponding parameter confidence intervals for $q_0$, $j_0$ and $s_0$ (through the Eqs. \ref{d1eq}, \ref{d2eq}, \ref{d3eq}) are:
$\left(
\begin{array}{ccc}
q_0 &  -1.09& 0.10 \\
 j_0 & -0.3  & 2.71 \\
s_0 &  -1.23716 & 8.34
\end{array}\label{cosmo}
\right)  $,
the best fit being $q_0=-0.51$, $j_0=0.69$ and $s_0=-0.48$, in agreement with other results in literature (see for instance \cite{vitagliano,gao}).
In Fig. (\ref{cosmogrb}) we plot our \textit{approximate cosmographic distance modulus} together with the SNIa Union dataset and the \textit{calibrated} GBRs HD. The reliability of the reconstruction is measured by the relative residuals $\epsilon= \frac{\mu_{obs} - \mu_{fit}}{\mu_{obs}}$, shown in Fig. \ref{res_cosmogrb}.
We finally perform our \textit{approximate cosmographic} analysis, considering a whole dataset containing both the SNIa Union dataset and the \textit{calibrated} GBRs HD, which we call the \textit{cosmographic dataset}.

We obtain the following parameter confidence ( at $2\sigma$) intervals for the $d_i$ and the cosmographic  parameters respectively:
$\left(
\begin{array}{ccc}

 d_1 &-1.35  & 2.26  \\
 d_2 & 0.71 &  1.30\\
 d_3& 0.39 & 0.59\,
\end{array}\label{cosmoall}
\right)$,\\

\begin{figure}
\includegraphics[width=7 cm]{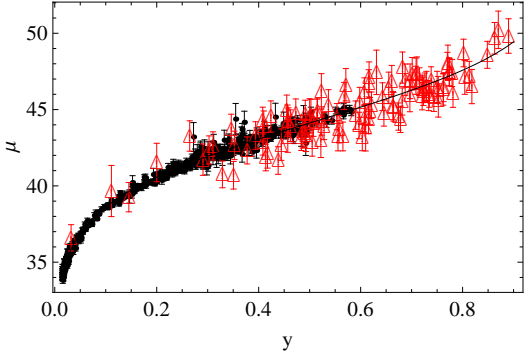}
\caption{Distance modulus for the best-fit values of our cosmography performed with the \textit{cosmographic dataset}. The filled black circles correspond to the SNIa Union dataset and the empty red triangles to the \textit{calibrated} GBRs HD.  }
\label{cosmogrball}
\end{figure}

\begin{figure}
\includegraphics[width=7 cm]{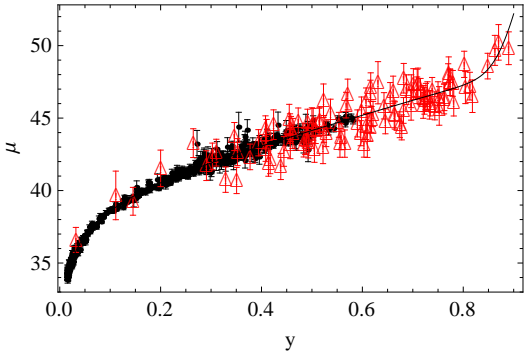}
\caption{Distance modulus for the best-fit values of the standard cosmography performed with the \textit{cosmographic dataset}. The filled black circles correspond to the SNIa Union dataset and the empty red triangles to the \textit{calibrated} GBRs HD. }
\label{cosmostandard}
\end{figure}

$\left(
\begin{array}{ccc}
q_0 &  -1.58 & -0.39 \\
 j_0 & 2.67  & 8.87 \\
s_0 & -6.88 & 46.9376
\end{array}
\right) $.
The best fit being $q_0=-0.9$, $j_0=5.25$ and $s_0=27.25$,  which again agree with the values found in literature. In Fig. \ref{cosmogrball} we show the \textit{approximate cosmographic distance modulus} together with the \textit{cosmographic dataset}. Moreover we observe that since the cosmographic dataset includes also the GBRs HD, which  spans a quite large range of redshift up to $z \simeq 8$,  the approximate function in Eq.(\ref{eq:approx}) is sampled also at higher values of the redshift, and its  behaviour  is not misshaped, even without any prior on the confidence regions, as nedeed above.
In order to further check the  reliability of our procedure, we compare the results summarized above
with that provided by the \textit{standard} cosmography. We apply this procedure to the  \textit{cosmographic dataset} only, and obtain the following parameter confidence ($2\sigma$) intervals:
$\left(
\begin{array}{ccc}
 q_0 &-1.09 & -0.08 \\
 j_0 & -0.30 & 2.71 \\
 s_0 & -1.24 & 8.34\,
\end{array}
\right)$,
which fully agree with the previous results. Also from the point of view of the analysis of residuals it turns out that our \textit{approximate cosmography } is statistically fully consistent with the \textit{standard cosmography}: actually we obtain a fully compatible values for the rootmean square  and the correlations in both cases. In Fig. \ref{cosmostandard} we show the \textit{standard cosmographic distance modulus} together with the \textit{cosmographic dataset}.
In short, let us note that we implemented a cosmographic analysis, using both the supernovae and the Gamma Ray Bursts data, which allow us to obtain  constraints on the parameters of cosmography (starting from our \textit{approximated luminosity distance}), which we also compare with the results of a \textit{standard} approach. Therefore our approach is very different from the one used in \cite{izzo}, where they obtain a cosmographic luminosity distance in the $y$-redshift, which is used to calibrate the $E_{\rm p,i}$ -- $E_{\rm iso}$ relation using a weighted $\chi^2$ estimator, but without making up the GRBs Hubble diagram. Moreover in that case the constraints on the cosmographic parameters are obtained from the SNIa data only.

\section{Discussion and Conclusions}
 Recently several interesting correlations among the Gamma Ray Burst
(GRB) observables have been identified. Proper evaluation and
calibration of these correlations are needed to use the GRBs as
standard candles to constrain the expansion history of the universe
up to redshifts of $z\sim 8$. Here we used the  GRB data set
recently compiled by Amati et al. (2008 - 2009)  to investigate, in a
quintessential cosmological scenario,  the $E_{\rm p,i}$ -- $E_{\rm iso}$ correlation relation. Marginalizing over the normalization, $b$, our Bayesian analysis provides the  following parameter confidence (at $3\sigma$) intervals for the the best-fit power-law and the intrinsic dispersion respectively:
$\left(
\begin{array}{ccc}
 a &1.263& 1.77 \\
 \sigma_{int}& 0.36 & 0.5,
\end{array}
\right)$.
The maximum likelihood value for the normalization coefficient turns out to be $b=52.67$. We used the fitted parameters to construct an \textit{ estimated fiducial} Hubble diagram that extends up to  redshifts $z\sim 8$. In the first part of our analysis we have assumed that the fit parameters do not change with the redshift. The limited number of GRBs prevents detailed exploration of the validity of this usually adopted working
hypothesis, which we tested somewhat investigating if the residuals  correlate with the redshift. We have not found any significant
correlation. Moreover we tested the fit of the $E_{\rm p,i}$ -- $E_{\rm iso}$ correlation parameters with respect to the evolution with redshift,
binning the GRBs into four groups with redshift from low to high, each group containing a reasonable number of GRBs.
We thus maximized the likelihood in each group of redshifts and determined the best fit calibration parameters. Our analysis indicates that the presently available GRBs datasets do not show \textit{statistically unambiguous} evolutionary effect with the cosmological redshift, but this should be tested further with larger GRBs samples.
We also investigated the impact of varying the parameters of our fiducial cosmological model, fitting the Amati
correlation relation on a regular grid in the space of parameters of our quintessential model, and evaluating, for each GRB, the root mean square of the percentage deviation from the fiducial $\mu$ value. In this way we performed a sort of average of the absolute percentage deviation which provides the variation of the fitting parameters $a$ and $b$. It turns out that the distance modulus may be under or overestimated by a modest $0.3\%$ with values never larger than $1\%$, so that the underlying cosmological model (in terms of varying the values of its characteristic parameters) has only a modest impact on the final estimate of the distance modulus.
However, this fact  does not imply that the cosmological
constraints that can be obtained from GRBs data with this method are also marginal: already in \cite{MEC10} we tested the stability of values of the fitted correlation parameters  by considering an ad hoc definition of the luminosity distance that gives much larger distances to objects at $z>2$ than either of the fiducial model. It turned out that such artificial \textit{crazy} luminosity distance is changing the values of the correlation parameters, but the difference is not dramatic at the $1 \sigma$ confidence level.
However the situation changes when we consider, in the space of parameters, the regions of  $1\,\sigma $, $2\,\sigma $ and $3 \,\sigma$ of confidence  for our \textit{crazy} model. It turns out that with respect to the same regions constructed for the fiducial model they overlap only at $ 1 \,\sigma$, but differ consistently at higher levels of confidence. As a consequence, when we made up the GRBs Hubble diagram the number of GRBs deviating from the fiducial $\mu$ more than $2\sigma$ increases from $N=14$, in the case of the quintessential cosmological model, to $N=51$ in the case of the \textit{crazy} model \footnote{In \cite{MEC10} the results concerning the \textit{crazy} model are inferred from fitting five correlation relations different from the Amati relation. However we tested that for the Amati relation also we obtain the same kind of behaviour.  }. The calibration of the Amati correlation, as well as the other known correlations, in several cosmological scenarios is therefore already needed to fully use the GRBs  as a cosmological probe. Actually, quite recently, in \cite{DOC11} the GRBs are used to test the $\Lambda$CDM vs. conformal gravity; and in a forthcoming paper we are standardizing the GRBs with  the Amati relation to test cosmological models based on extended theories of gravity. 
 Moreover we apply a local regression technique to estimate, in a model independent way, the distance modulus from the recently updated SNIa sample. The derived calibration parameters are used to construct an updated GRBs Hubble diagram, which we call the \textit{calibrated} GRBs HD. We also compare the  \textit{fiducial} and \textit{calibrated} GRBs HDs, which turned out to be  fully statistically consistent, thus indicating that they are not affected by any systematic bias induced by the different calibration procedures. This means that the high redshift GRBs can be used to test different models of dark energy settling the circularity problem.
Finally we propose a new approach to calibrate the GRB relations in a  \textit{cosmologically independent} way, by using an approximate
luminosity distance relation, which holds in any cosmological model.
We use this calibration of the Amati relation to construct an \textit{empirical approximate}  HD, which is fully consistent with the \textit{calibrated} GRBs HD, probing in such a way the reliability of our approach, which could provide a robust procedure to calibrate in a cosmological independent way different GRBs correlation relations, specially if the available dataset are poorly populated in the SNIa range of redshift, so that, in such a case, the SNIa fitting procedure fails. We finally investigated the implications of such an approach for the high redshift \textit{cosmography}. Actually, starting from the estimation of the  constant $d_1$, $d_2$ and $d_3$ present in our approximate luminosity distance relation, we constructed the map which connects our $d_i$s to the parameters describing the kinematical state  of the universe $q_0$, jerk $j_0$, and snap $s_0$. This map is a \textit{core} for our \textit{approximate cosmography }, which we actually applied to a whole dataset containing both the SNIa Union dataset and the \textit{calibrated} GBRs HD, which we call the \textit{cosmographic dataset}. Using the \textit{cosmographic dataset}, we show that the deceleration parameter $q_0$ up to the $3 \sigma$ confidence level is definitively negative. The constraints for the jerk and snap parameters are, instead, less strong.
Finally it turns out that our results are fully consistent with the results obtained using the standard approach, showing that our new  \textit{approximate cosmography } allows robust results in a wide range of redshifts.
\subsection*{Acknowledgments}
We warmly thank Dr Lorenzo Amati for providing us with data of 14 unpublished GRBs, which we added to the samples in (Amati et al. 2008) and (Amati et al. 2009).
This paper was supported in part by the Polish Ministry of Science and Higher Education grant NN202-091839.

\end{document}